\documentclass[aps,prl,twocolumn]{revtex4}
\usepackage{graphicx}
\usepackage{amssymb}
\newcommand{ \be}{\begin{equation}}
\newcommand{ \ee}{\end{equation}}
\newcommand{\beq}{\begin{eqnarray}}
\newcommand{\eeq}{\end{eqnarray}}
\newcommand{\bem}{\begin{pmatrix}}
\newcommand{\eem}{\end{pmatrix}}
\newcommand{\bmx}{\begin{array}}
\newcommand{\emx}{\end{array}}

\begin{document}

\title{Structural properties of dense hard sphere packings}

\author{B.A. Klumov$^{1,2}$, S.A. Khrapak$^{1,2}$,
G.E. Morfill$^{1}$}
\affiliation{$^1$Max-Planck-Institut f\"ur extraterrestrische Physik, D-85741 Garching,
Germany \\$^2$Joint Institute for High Temperatures, 125412 Moscow, Russia}
\date{\today}

\begin{abstract}
The structural properties of dense random packings of identical hard spheres (HS) are investigated. The bond order parameter method is used to obtain detailed information on the local structural properties of the system for different packing fractions $\phi$, in the range between $\phi=0.53$ and $\phi=0.72$. A new order parameter, based on the cumulative properties of spheres distribution over the rotational invariant $w_6$, is proposed to characterize crystallization of randomly packed HS systems. It is shown that an increase in the packing fraction of the crystallized HS system first results in the transformation of the individual crystalline clusters into the global three-dimensional crystalline structure, which, upon further densification, transforms into alternating planar layers formed by different lattice types.
\end{abstract}

\pacs{{\bf 45.70.-n, 61.50.Ah, 05.20.Jj}}
\maketitle

The model of hard spheres (HS) is of fundamental importance in condensed matter physics and material science since it successfully reproduces the essential structural properties of liquids, crystals, glasses, colloidal suspensions and granular media. Packings of HS have also been used in solving important problems of information and optimization theories \cite{parisi,still}. In this Letter
we focus on the structural properties of dense three dimensional (3D) HS systems at different packing fractions $\phi=\frac{\pi}{6}\rho\sigma^3$, where $\rho=N/V$ is the density of $N$ hard spheres in a system volume $V$ and $\sigma$ is the diameter of the spheres. In particular, we take a large set of packings composed of $N=10^4$ identical spheres with periodic boundary conditions,
generated using Jodrey-Tory (JT) \cite{jta,jtb} and Lubashevsky-Stillinger (LS) \cite{ls} algorithms as described in detail in Refs.~\cite{anik1,anik2}. The corresponding packing fractions vary in the ranges $\phi \simeq 0.53 - 0.71$ and $\phi = 0.58-0.68$ for JT and LS protocols, respectively. Both ranges include the random close packing state (RCP) at $\phi_{\rm c}\simeq 0.64$ (Bernal limit) \cite{bernal}.
The main purpose is to look into the details on how densification of the disordered solid state affects the local and global order of the HS system.

\begin{figure}
\includegraphics[width=8.4cm]{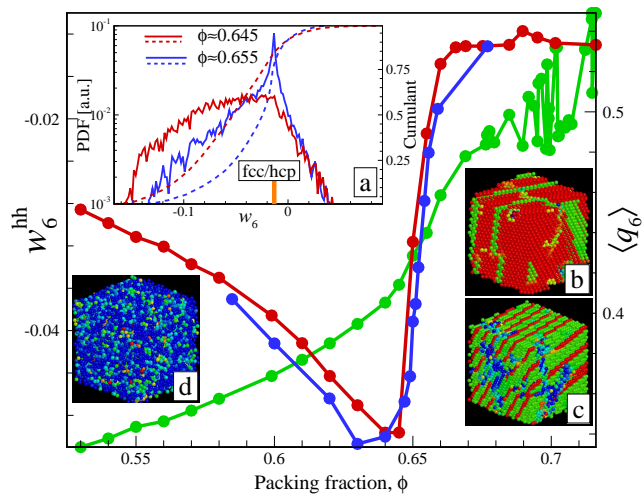}
\caption{(Color) Order parameter $w_6^{\rm hh}$ (red and blue lines correspond to
JT and LS packing protocols, respectively) and the mean value of the rotational invariant $\langle q_6 \rangle$ (green) versus packing fraction $\phi$ for the dense randomly packed HS system. Explosive-like growth of the parameter $w_6^{\rm hh}(\phi)$ at $\phi \gtrsim \phi_{\rm c}\simeq 0.64$ is a signature of the appearance of crystalline order (hcp-like and fcc-like spheres) in the system. On the other hand, increase in $\langle q_6 \rangle$ near $\phi_{\rm c}$ is monotonous and slow. Inset (a) shows distributions of spheres over the parameter $w_6$ for the two packing fractions ($\phi \approx 0.655$,  blue solid curve; and $\phi \approx 0.645$,  red solid curve). Cumulants of these distributions are plotted by dashed curves of the corresponding color. Emergence of the peak in the distribution over $w_6$ corresponds to the onset of crystallization in the system. The values of $w_6$ for perfect hcp and fcc lattices are also indicated ($w_6^{\rm hcp} \simeq -1.25 \times 10^{-2}$, $w_6^{\rm fcc} \simeq -1.31 \times 10^{-2}$). Insets (b, c) demonstrate two realizations of sphere packing with very close packing fractions near $\phi \approx 0.7$, which have quite different  $\langle q_6 \rangle$ values. Example of disordered packing at $\phi \approx 0.55$ is shown in the inset (d). The spheres are color-coded by their $q_6$ value (red and green colors correspond to fcc- and hcp-like spheres, respectively), revealing fcc-dominated (b) and hcp-dominated (c) packings. The possibility for the dense HS systems to have quite different composition (with different relative density of fcc and hcp crystalline configurations) is the main reason behind the pronounceable oscillations in $\langle q_6 \rangle$ observed for $\phi \gtrsim 0.68$.}
\label{op}
\end{figure}

\begin{figure}
\includegraphics[width=8.4cm]{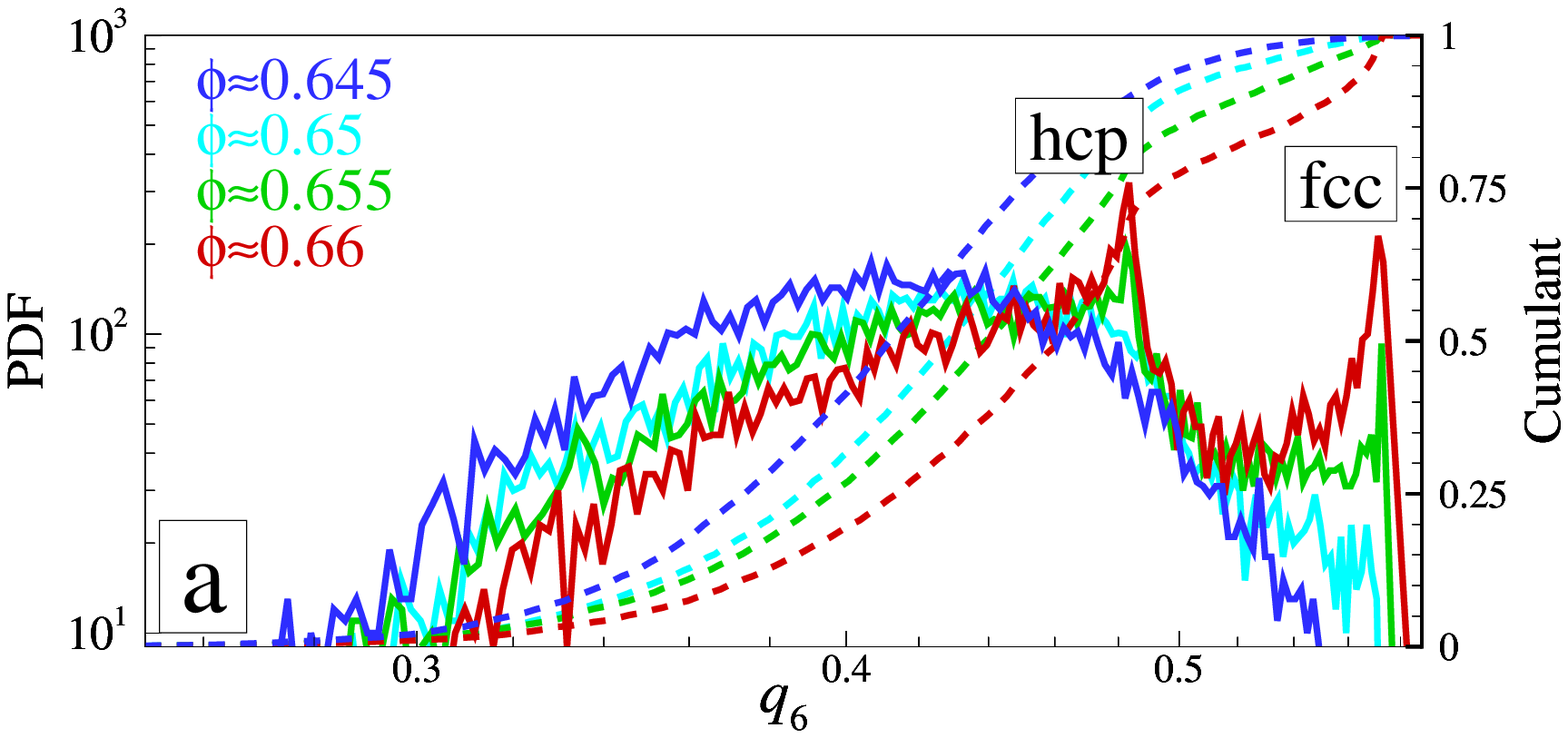}\\
\hspace{-0.7cm}
\includegraphics[width=7.5cm]{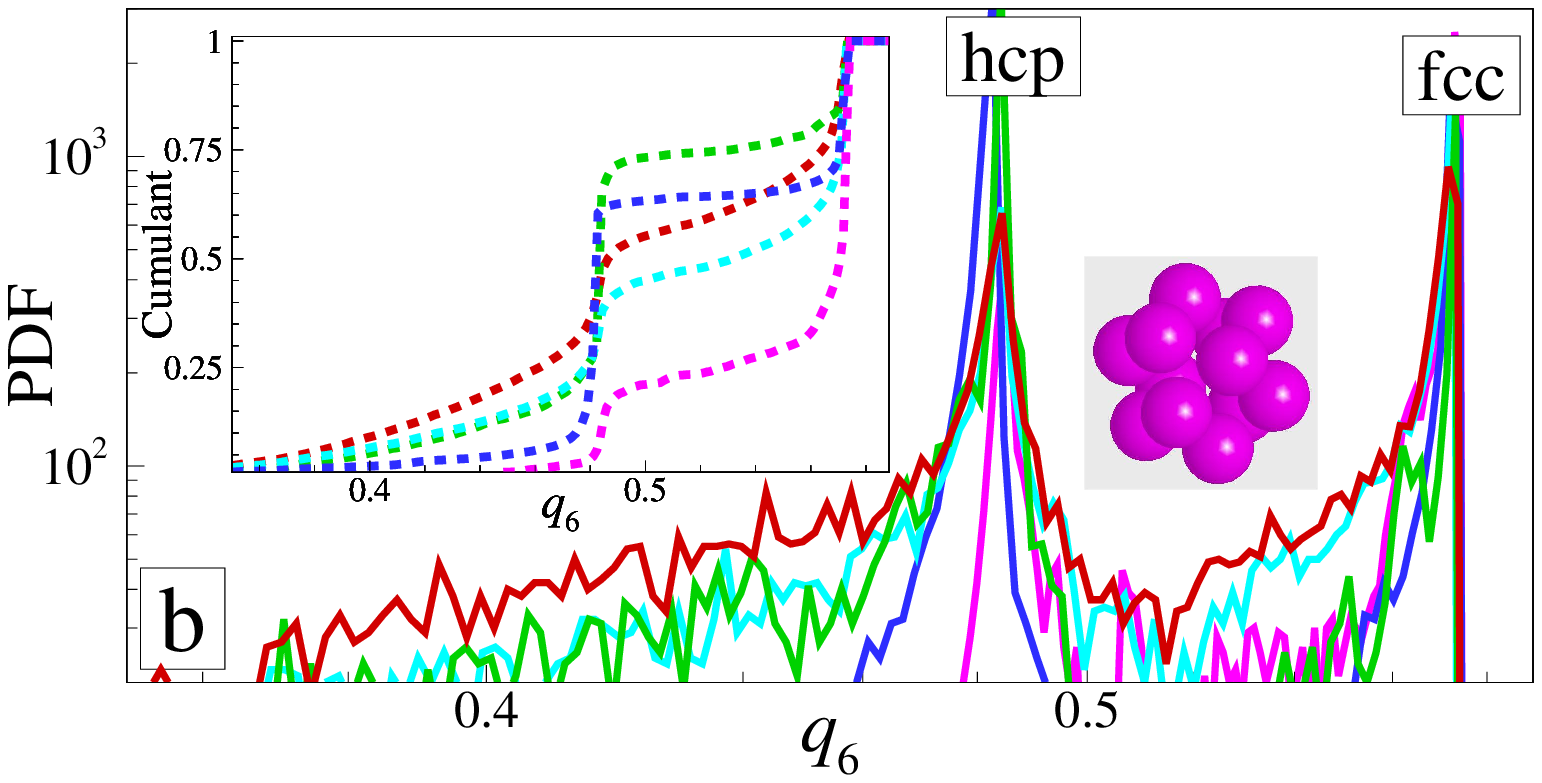}
\caption{(Color). Probability distribution functions (PDFs) of the rotational invariant $q_6$ for different packing fractions $\phi$ in the regimes corresponding to the onset of crystallization (a) and to the dense crystallized HS systems (b). Top panel (a) shows the PDFs for $\phi \simeq 0.645$ (blue), $\phi \simeq 0.65$ (cyan), $\phi \simeq 0.655$ (green) and $\phi \simeq 0.66$ (red). The values of $q_6$ for perfect hcp and fcc lattices are also indicated ($q_6^{\rm hcp} \simeq 0.48$, $q_6^{\rm fcc} \simeq 0.57$). Bottom panel (b) presents the PDFs for $\phi \simeq 0.685$ (red), $\phi \simeq 0.695$ (green), $\phi \simeq 0.71$ (cyan), $\phi \simeq 0.715$ (blue) and $\phi \simeq 0.716$ (magenta). In the latter case quasi-crystalline (QC) spheres with $q_6^{\rm QC} \simeq 0.52$ ($q_6^{\rm hcp} <q_6^{\rm QC} < q_6^{\rm fcc}$) can be identified (the typical QC arrangement is shown in the bottom panel; the cluster looks like torsional modification of hcp/fcc lattices \cite{3Da}). The cumulants  $Q_6$ of these distributions, $Q_6(x) \equiv \int_{-\infty}^x n(q_6)dq_6$ are plotted by dashed lines of the same colors. These curves, in particular, demonstrate abundance of the hcp- and fcc-like spheres and reveal the oscillatory behavior of $n_{\rm fcc}$ and $n_{\rm hcp}$ for sufficiently dense packing (cf. variations of $\langle q_6 \rangle$ for $\phi \ge 0.68$ in Fig.~\ref{op}).}
\label{q6}
\end{figure}

To define the local structural properties of the system we use the bond order parameter method \cite{stein}, which has been widely used in the context of condensed matter physics~\cite{stein,torqa},
HS systems~\cite{troadec,valera,volkov,torqb,aste,loch,medved1,li,medved2,makse}, complex plasmas~\cite{nat,3Da,mitic,3Db,khr}, colloidal suspensions~\cite{coll0,auer,solomon,coll1,coll2}, granular media~\cite{gran1}, etc. In this method the rotational invariants of rank $l$ of both second $q_l(i)$ and third $w_l(i)$ order are calculated for each sphere $i$ in the system from the vectors (bonds) connecting its center with the centers of its $N_{\rm nn}(i)$ nearest neighboring spheres:
\be
q_l(i) = \left ( {4 \pi \over (2l+1)} \sum_{m=-l}^{m=l} \vert~q_{lm}(i)\vert^{2}\right )^{1/2}
\ee
\be
w_l(i) = \hspace{-0.8cm} \sum\limits_{\bmx {cc} _{m_1,m_2,m_3} \\_{ m_1+m_2+m_3=0} \emx} \hspace{-0.8cm} \left [ \bmx {ccc} l&l&l \\
m_1&m_2&m_3 \emx \right] q_{lm_1}(i) q_{lm_2}(i) q_{lm_3}(i),
\label{wig}
\ee
\noindent
where $q_{lm}(i) = N_{\rm nn}(i)^{-1} \sum_{j=1}^{N_{\rm nn}(i)} Y_{lm}({\bf r}_{ij} )$, $Y_{lm}$ are the spherical harmonics and ${\bf r}_{ij} = {\bf r}_i - {\bf r}_j$ are vectors connecting centers of spheres $i$ and $j$. In Eq.(\ref{wig}) $\left [ \bmx {ccc} l&l&l \\ m_1&m_2&m_3 \emx \right ]$ are the Wigner 3$j$-symbols, and the summation in the latter expression is performed over all the
indexes $m_i =-l,...,l$ satisfying the condition $m_1+m_2+m_3=0$. In 3D the densest possible packings of identical hard spheres are known to be face centered cubic (fcc) and hexagonal close-packed (hcp) lattices ($\phi \simeq 0.74$). To detect these structures we calculate the rotational invariants $q_4$, $q_6$, and $w_6$ for each sphere using the fixed number of nearest neighbors  $N_{\rm nn}=12$. A sphere whose coordinates on the plane $(q_4, q_6)$ are sufficiently close to those of the ideal fcc (hcp) lattice is counted as fcc-like (hcp-like) sphere.

\begin{figure}
\includegraphics[width=8.4cm]{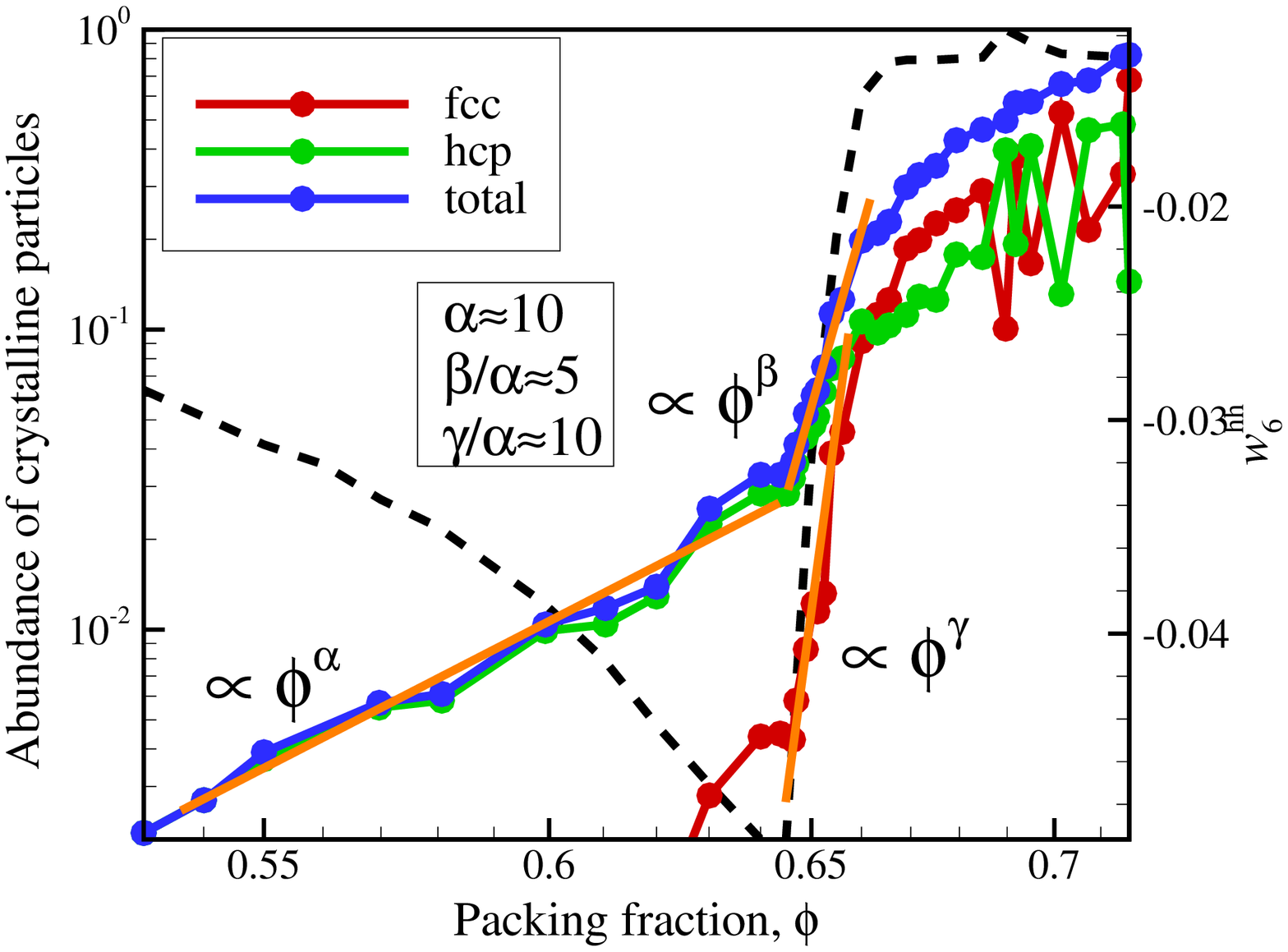}
\caption{(Color). Relative density of hcp-like and fcc-like spheres, $n_{\rm hcp}$ (green) and $n_{\rm fcc}$ (red), as well as the total density of crystalline spheres $n_{\rm tot}\equiv n_{\rm hcp}+n_{\rm fcc}$ (blue), versus the packing fraction $\phi$. For moderate packing fractions ($\phi \le 0.645$), the increase of $n_{\rm tot}$ with $\phi$ is due to the appearance of single hcp-like spheres.
The increase has a power-law character ($n_{\rm tot} \propto \phi^{\alpha}$, where $\alpha \simeq 10$). Much more rapid power-law growth of both $n_{\rm tot}$ and $n_{\rm fcc}$ takes place for $\phi \gtrsim 0.645$ ($n_{\rm tot} \propto \phi^{\beta}$, $n_{\rm fcc} \propto \phi^{\gamma}$, where $\beta \simeq 50$ and $\gamma \simeq 100$). Transition from hcp- to fcc-dominated packing ($n_{\rm hcp} \lesssim n_{\rm fcc}$) occurs at $\phi \simeq 0.66$. Oscillations of $n_{\rm fcc}$ and $n_{\rm hcp}$ which are clearly seen for sufficiently dense systems ($\phi \gtrsim 0.68$) reflect two preferable possibilities of HS packing (cf. Figs.~\ref{op} and~\ref{q6}). The dependence of the order parameter $w_6^{\rm hh}$ on $\phi$ is also plotted (dashed black line) to indicate the onset of crystallization.}
\label{solid}
\end{figure}

\begin{figure}
\includegraphics[width=8.4cm]{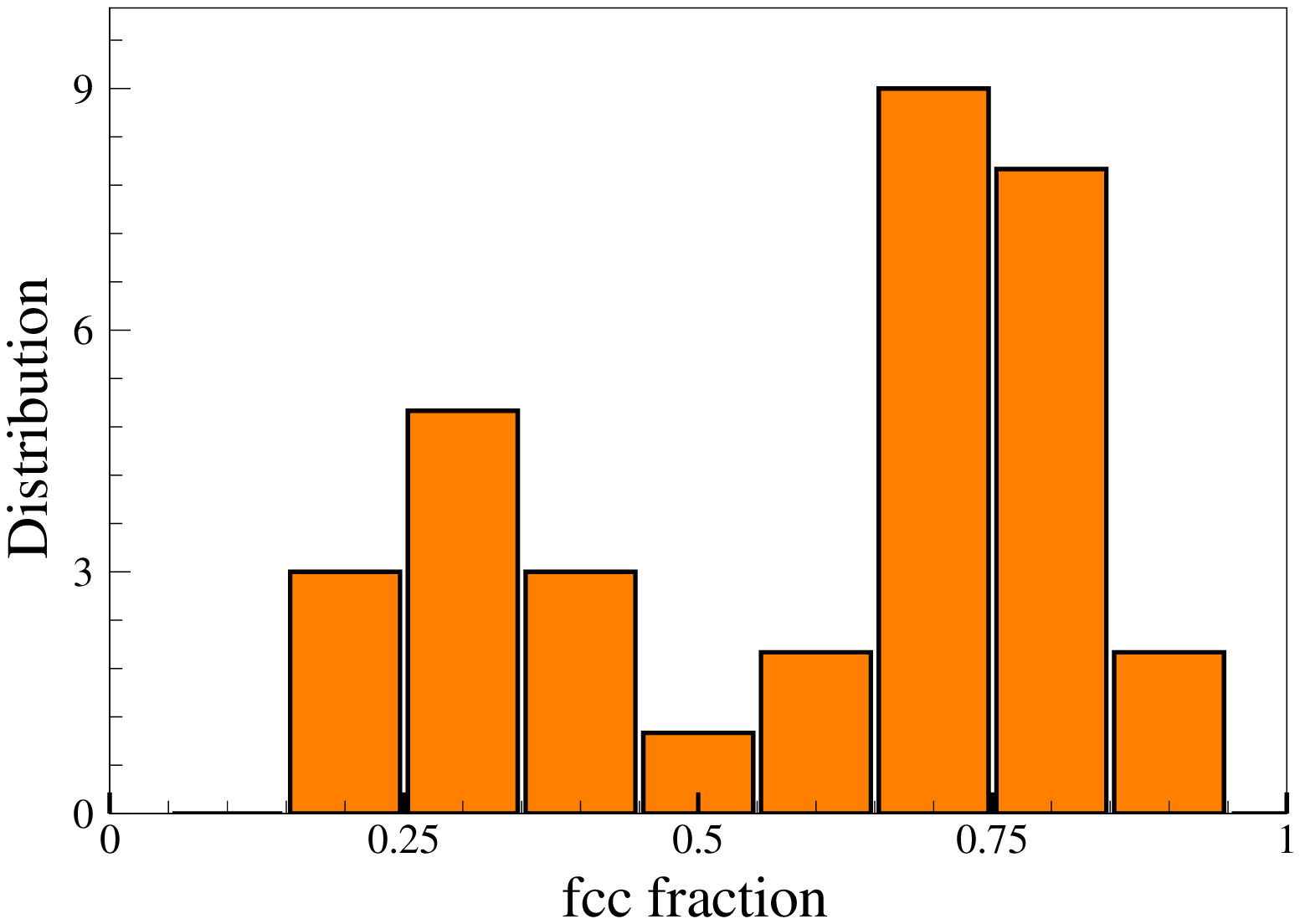}
\caption{(Color online). Relative fraction of fcc-like spheres, $n_{\rm fcc}/(n_{\rm fcc} + n_{\rm hcp})$, for dense HS packings ($\phi \ge 0.68$). About 35 realizations generated by the JT algorithm are analyzed. Two peaks reflect domination of either fcc or hcp phases in the system.}
\label{hist}
\end{figure}

It has been shown recently \cite{melt_freez} that the cumulative distribution function of the form
\be
\hat W_6(x) \equiv \int_{-\infty}^x n(w_6)dw_6
\ee
is an extremely sensitive measure of the structural order in the system. Here $n(w_6)$ is the distribution of spheres over the rotational invariant $w_6$ normalized to unity. The relevant order parameter is the position of the half-height of the cumulative distribution $w_6^{\rm hh}$, so that $\hat W_6(w_6^{\rm hh}) =1/2$. Let us check whether this order parameter can be used to identify the onset of crystallization in the randomly packed HS system.

Figure~\ref{op} shows the dependence of the order parameter $w_6^{\rm hh}$ and the mean value of $q_6$ [defined as $\langle q_6 \rangle = (1/N)\sum\limits_{i=1}\limits^N q_6(i)$] on the packing fraction $\phi$. Both protocols used here (JT and LS) reveal nearly identical structural properties as clearly seen from Fig.~\ref{op}. Explosive-like growth of the order parameter $w_6^{\rm hh}$ at $\phi>\phi_{\rm c}$ reflects the appearance of crystalline order (hcp-like and fcc-like spheres). This proves the extreme usefulness of this structural measure to locate the onset of crystallization, as demonstrated here in the context of randomly packed HS systems. Inset (a) shows probability distribution functions (PDFs) of $w_6$ for the two packing fractions [$\phi \simeq 0.645$ (red curve) and $\phi \simeq 0.655$ (blue curve)]. The peak on the presented PDF corresponding to the denser system (blue curve) corresponds to the appearance of hcp-like spheres. Cumulative distributions for these PDFs are also plotted making clear the basic idea: It is very convenient to employ smooth cumulative distributions of this kind to characterize structural properties of systems near crystallization. The dependence of the mean $q_6$ value on $\phi$ shown in Fig.~\ref{op} also demonstrates interesting properties. Relatively
slow growth of $\langle q_6 \rangle$ for $\phi \lesssim \phi_{\rm c}$ is associated with the appearance of single hcp-like spheres \cite{medved2}. For $\phi> \phi_{\rm c}$ the parameter $\langle q_6 \rangle$ grows more rapidly. Further densification of the system results in strong oscillations of $\langle q_6 \rangle$, which start at $\phi \simeq 0.68$. Geometry gives a simple explanation of such an oscillatory behavior: The point is that the hcp and fcc crystal structures are characterized by quite different values of the parameter $q_6$ (for ideal  hcp and fcc lattices $q_6^{\rm hcp} \approx 0.48$ and $q_6^{\rm fcc} \approx 0.57$). Dense HS packings can be apparently realized (for a given $\phi$) with significantly different relative quantity of hcp and fcc phases [see insets (b) and (c) in Fig.~\ref{op}, demonstrating hcp-dominated and fcc-dominated packings, respectively). This explains the oscillations. Note that similar results were obtained recently in \cite{medved2, makse}. Returning to the regime of packing fractions near $\phi_{\rm c}$, Fig.~\ref{op} clearly demonstrate that the order parameter $w_6^{\rm hh}$ is considerably more sensitive to the onset of crystallization in the HS system as compared to the mean value of the rotational invariant $\langle q_6 \rangle$.

Figure~\ref{q6} shows the probability distribution function of the order parameter $q_6$ for different packing fractions corresponding to the onset of crystallization (top panel) and dense crystallized packings (bottom panel). Cumulative distributions $Q_6(x) \equiv \int_{-\infty}^x n(q_6)dq_6$ are also shown for each PDF. The relative numbers of spheres in hcp-like ($n_{\rm hcp}$) and fcc-like ($n_{\rm fcc}$) state are plotted in Fig.~\ref{solid} as functions of $\phi$. The total number of spheres involved in crystalline arrangements ($n_{\rm tot}=n_{\rm hcp}+n_{\rm fcc}$) is also shown. For $\phi \lesssim \phi_{\rm c}$ the relatively slow increase of $n_{\rm tot}$ with $\phi$ is associated with emergence of single hcp-like spheres, which are nearly randomly distribute over the system volume. The increase has a power-like character ($n_{\rm tot} \propto \phi^{\alpha}$ with $\alpha \simeq 10$). At $\phi\simeq \phi_{\rm c}$ the slopes of $n_{\rm tot}(\phi)$ and $n_{\rm fcc}(\phi)$ drastically increase. The number of fcc-like spheres reaches that of the hcp-like spheres ($n_{\rm fcc} \approx n_{\rm hcp}$) at $\phi \simeq 0.66$, where transition from hcp-dominated to fcc-dominated packing occurs. Oscillations of $n_{\rm fcc}$ and $n_{\rm hcp}$, clearly seen for $\phi \gtrsim 0.68$ correspond to the property of dense HS packings discussed above (see Figs.~\ref{op} and \ref{q6}) and also observed in Ref.~\cite{medved2} with the help of Delaunay simplexes.

Figure~\ref{hist} shows the probability distribution of the ratio $n_{\rm fcc}/(n_{\rm fcc} + n_{\rm hcp})$ for the dense ($\phi \ge 0.68$) HS systems. Two peaks of the distribution indicate that dense HS systems prefer to be dominated by either fcc phase (with somewhat higher probability) or by hcp phase (with somewhat lower probability). Equal distributions between fcc and hcp phases are considerably less probable. However, the lack of statistics (only 35 packings were available for the analysis) and specifics of the JT algorithm may also affect this observation. For this reason we do not elaborate further on this point here.

\begin{figure}
\includegraphics[width=8.4cm]{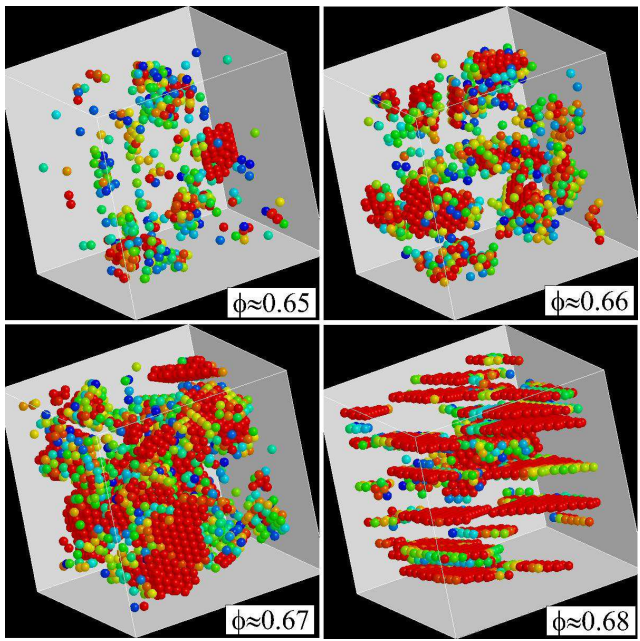}
\caption{(Color). Distribution of fcc-like arrangements for four different packing fractions $\phi$ of the dense HS system. Spheres are color-coded by the value of the rotational invariant $q_6$ (red color corresponds to fcc-like spheres). At low $\phi$ only a few individual fcc-like spheres and small clusters from fcc-like spheres can be seen. Densification leads to the formation of large 3D clusters of fcc-like spheres. Eventually, clusters of fcc-like and hcp-like spheres transform into a global crystalline structure. Further densification results in the structural transition, when essentially 3D shape of this global crystalline structure suddenly changes to planar 2D layers formed by fcc-like (shown in the figure) and hcp-like spheres.}
\label{cluster}
\end{figure}

Cluster analysis of the arrangements of hcp-like and fcc-like spheres reveals another interesting structural property of dense HS packings. Figure~\ref{cluster} shows how the distribution (and the shape) of clusters composed of fcc-like spheres varies upon densification of the system. For relatively low volume fractions, $\phi \lesssim 0.65$, fcc-like clusters consist of only few spheres. Increase of $\phi$ results in the appearance of several big (containing few tens of spheres) fcc-like clusters. A typical cluster has a complicated 3D shape (see Fig.~\ref{cluster} for $\phi\simeq 0.66$). Eventually, nuclei of fcc-like and hcp-like clusters transform into global crystalline structure occupying most of the system volume. Further increase in $\phi$ leads to a structural change: a transition from essentially 3D character of this structure to 2D layers composed of fcc-like and hcp-like spheres occurs. This transition is illustrated in Fig.~\ref{cluster} (for $\phi\simeq 0.68$). An example of such ``sandwich-like'' structure can be also seen in Fig.~\ref{op}(c).

To conclude, we have demonstrated that the order parameter, based on the cumulative distribution of spheres over the values of the rotational invariant $w_6$, is a very sensitive measure of the onset of crystallization in the system of randomly packed hard spheres. The proposed order parameter is especially convenient indicator of the appearance of hcp and fcc crystallites in the system. We used this to investigate how the relative distribution between  hcp and fcc arrangements varies with increasing the sphere packing fraction. We observed that at  $\phi \simeq 0.68$ structural transition occurs: The essentially three dimensional shape of the global crystalline aggregate suddenly changes to planar sandwich-like (layered) structure.

\acknowledgements
We wish to acknowledge with deep appreciation and gratitude the extremely valuable discussions
with N.N. Medvedev. We thank A.V. Anikeenko for providing us with HS data sets. This work was partly supported by DLR under Grant 50WP0203. (Gef\"{o}rdert von der Raumfahrt-Agentur des Deutschen Zentrums f\"{u}r Luft und Raumfahrt e. V. mit Mitteln des Bundesministeriums f\"{u}r Wirtschaft und Technologie aufgrund eines Beschlusses des Deutschen Bundestages unter dem F\"{o}rderkennzeichen 50 WP 0203.)

\end{document}